\documentclass[10pt,a4paper]{article}
\usepackage{amsmath,amssymb}
\usepackage{url}
\usepackage{placeins}
\usepackage{subfigure}
\usepackage{multirow}
\usepackage{epsfig}

\newcommand\ben{\begin{equation}}
\newcommand\een{\end{equation}}
\newcommand\bea{\begin{eqnarray*}}
\newcommand\eea{\end{eqnarray*}}
\newcommand\bean{\begin{align}}
\newcommand\eean{\end{align}}

\author{Chris Kenyon and Andrew Green\footnote{Contact: chris.kenyon@lloydsbanking.com}}
\title{Will Central Counterparties become\\ the New Rating Agencies?\footnote{\bf The views expressed are those of the authors only, no other representation should be attributed.}
}
\date{\today\\ \vskip5mm Version 2, draft 1.00}

\begin{document}

\maketitle

Central Counterparties (CCPs) are widely promoted as a requirement for safe banking  \cite{BCBS-189,BCBS-206,BCBS-219}, with little dissent except on technical grounds (such as proliferation of CCPs) \cite{Duffie2011a,Singh2012a}.  Whilst CCPs can have major operational positives, we argue that CCPs have many of the  business characteristics of Rating Agencies, and face similar business pressures.  Thus we see a risk that prices from CCPs may develop the characteristics attributed to ratings from Rating Agency pre-crisis \cite{Tomlinson2007a,Benmelech2009a,FCEC2011a}.  Business over-reliance on ratings of questionable accuracy is seen as a cause of the financial crisis \cite{Partenoy2008a,FCEC2011a}.  We see the potential for same situation to be repeated with prices from CCPs.  Thus the regulatory emphasis on CCPs, rather than on collateralization, may create the preconditions for an avoidable repeat of the financial crisis.

The outline of the paper is as follows: we first describe the functions of a CCP, highlighting how they are carried out; then we give CCP business characteristics, comparing and contrasting them with Rating Agencies; next we develop the potential consequences.  We conclude by emphasising the key risk mechanism (complacency with CCP prices) and our recommendation (avoid privileging any price source or method).

\subsection*{CCP Functions}

The main function of a CCP is to remove credit risk transmission by intermediating trades and by only doing collateralized trades.  A CCP does not remove credit risk.  CCPs only block {\it one} credit risk transmission route: direct counterparty exposure.  If there is a systematic crisis, e.g. assets dropping suddenly in value, CCPs offer no protection.  The issue in a systematic asset price crisis is not counterparty exposure but credit risk from asset value (or rather, the lack of it).  Indeed since CCPs transform credit risk transmission into liquidity risk \cite{Gregory2012a} and since they actualize market volatility on a day-to-day basis, CCPs may actually increase default risks \cite{Kenyon2012a}.

A CCP must set prices on all its trades in order to call, or return, collateral.  In general CCPs base prices as far as possible on transaction prices from the channels that they work with (exchanges, ECNs, brokers).  Does this means that CCP prices are correct?  By definition a transaction price is correct {\it for that transaction} because that is where two counterparties were willing to do business {\it at that point in time}.  This does not mean that those prices have significance beyond the actual transaction; nor does it mean that the prices are stable over time.  All CDO tranche prices that were transacted were correct (by definition) at transaction time.  Their future value was highly variable.

A CCP may use and publish pricing methods, or pricing principles so its counterparties can reproduce its prices.  The CME has stated that it prices swap futures using a patented method licensed from Goldman Sachs \cite{Cameron2012a}.  The fact that a CCP uses a pricing method does not mean that it is correct in the sense of having a longterm validity, for example LCH changed from Libor discounting to OIS discounting for its portfolio of swaps on 29 June 2010 for USD, EUR and GBP \cite{LCH2010a}.   Pricing methods change.  Clarity of pricing method, whatever the level of industry use, is no guarantee of long term validity, as the CDO crisis made painfully obvious.  This does not mean that the principle of no-arbitrage is violated, just that its application depends on the circumstances, and these change (assuming that it is used). 

CCPs encourage professional handling of collateral by requiring collateralized trades, this is a major operational positive.  The Lehman default, and other defaults, showed just how badly collateral can be managed by financial institutions \cite{Carlsson2010a,Kary2012a}.  It might be imagined, given the vast sums involved, that collateral would be treated with care.  However, from a profit perspective --- historically --- there has been little motivation.  Today it is becoming evident that collateral management will form a key part of profitability and has significant effects on pricing \cite{Fuji2011a,Piterbarg2012a}.  

\subsection*{Business Characteristics}

We start from Rating Agencies and see where their business characteristics have similarities with CCPs, either in general, or pre-crisis.

Rating Agencies form an oligopoly with high barriers to entry.  CCPs are in a similar situation.  It is hard to set up a new CCP: regulators must provide recognition; channels and direct clients must be persuaded why this new CCP is better than existing CCPs, they must set up new (and expensive) physical and logical connections.  If national CCPs become commonplace with Basel III then the oligopoly situation will be repeated nationally.

Ratings are widely used and have a privileged position in many contexts: firm articles of associations (e.g. permitted investments); law (e.g. permitted investments by government bodies); regulations (e.g. determining capital requirements).  Similarly prices from CCPs are privileged in that any firm wanting to have the regulatory benefit of using a CCP (lower capital) must accept the prices from the CCP.  It is also envisaged that by law, in some jurisdictions, some trade types will have to be traded through a CCP.  In this situation counterparties have no choice but to accept CCP prices.  Thus CCP pricing is on track to gain a level of widespread use and privilege similar to Ratings.

Pre-crisis Rating Agencies faced pressure to increase their product coverage because of the privileges of official ratings, and did so.  This gained Rating Agencies significant extra revenue.  In 2006 Moody's reported that Structured finance (largely CDOs) accounted for 44\%\ of their income, for Fitch the proportion was 51\%\ \cite{Tomlinson2007a}. Given the regulatory benefits to clients of using CCPs, we see CCPs facing the same business pressure from clients (and regulators) to increase product coverage.  

Rating Agencies supply a good (ratings) whose quality has little direct effect on them.  If a rating suddenly has to be changed by a large amount it is not the Rating Agency that suffers a direct consequence (apart from reputation).  As part of the CCP function they supply a good (prices) whose quality also has little direct effect on them.  All CCP trades are back to back, collateralized, and have Initial Margins.  If the prices, or pricing methodologies, are found to require significant widespread change, CCPs have little direct risk.  

Rating Agencies face the usual business pressures of competition, and hence the pressure to increase revenue and/or reduce costs.  Competition between CCPs is already being stoked by cost comparisons between them \cite{Shah2012a}.  

\subsection*{Consequences}

The two most significant potential consequences of the similarities in business characteristics between CCPs and Rating Agencies are linked: 
\begin{itemize}
	\item an assumed safety from having CCPs may produce a complacency that CCP prices and pricing methods are correct in terms of having long term validity or stability;
	\item the privileged status of CCP prices may grow over time, for example we can imagine Asset Managers being {\it required} by law to use CCP prices internally.
\end{itemize}
These consequences essentially concern quality and reliability of prices and pricing methods, how they are regarded in general, in regulations, and legally.  

Whilst CCPs keep to vanilla products then these risks may be limited, although they will still be present as the Libor to OIS example above demonstrates.  However, as competitive pressures move CCPs towards new and more complex products, as Rating Agencies moved into more complex and profitable products, these risks will increase.  We can expect a wave of new products to devised in response to regulations, for example to mitigate capital requirements.  This has already started with the introduction of swap futures on the CME and so the potential risks are already appearing.

What were the consequences of perceived poor quality ratings pre-crisis on Rating Agencies?  No Rating Agency went bankrupt.  It is only now, four years later, that there has been a single significant legal judgment against Rating Agency practices pre-crisis \cite{Jagot2012a}.  What lessons might a CCP draw from this history?

Where an oligopoly delivers services that are difficult to reproduce, authorities may be reluctant to reduce that number, whatever the quality of those services.  If they reduce the number by one, as effectively happened after the Enron and Worldcom accounting events, then the reluctance to act again may be significantly higher.  This appears analogous to the situation with Rating Agencies: not too-big-to-fail but perhaps too-difficult-to-replace. 

Pre-crisis how could a deal team get the rating that they wanted for a given deal?  There is increasing evidence that shopping around for favorable ratings was widespread \cite{Benmelech2009a,Griffin2012a}, just apart from deal teams working closely with Rating Agencies to achieve a particular rating \cite{Tomlinson2007a,Jagot2012a}.  We can similarly imagine major customers working closely with CCPs on pricing methodologies, which may already be happening \cite{Cameron2012a}.   Alternatively, clients can move to the CCP that has prices that are favorable to them.

Since CCPs face normal business pressures they will try to decrease their costs.  Quants are expensive so we can envisage significant pressure to have the minimum required.  Traditionally banks have been perceived to pay quants more than service providers (however important) so it may be difficult for CCPs to hold on to their most skilled staff.  This produces a secondary issue, when clients dispute prices or methodologies with the CCP, the CCP will be under double pressure to agree: firstly the clients may go elsewhere, and secondly the client quants may be perceived to be more skilled (although clearly not disinterested).

How much confidence in institutional prices is reasonable?  Those capable of finding out that the correct price is different have no motivation to publicize the fact.  They can simply trade in the favorable direction. Alternatively, they can take positions and then publicize the correct pricing method, aiming to profit as the market standard changes.  There are limits to this arbitrage because CCPs typically require increasing IM with increasing exposure which at some point would render such trades uneconomic.  However, limits do not mean that it will not happen.

\subsection*{Conclusions}

If there are privileged prices from institutions then the pre-conditions for a new financial crisis based on over-reliance on those prices are in place.  This would be an avoidable re-run of the perceived contribution of over-reliance of official ratings to the previous financial crisis \cite{Partenoy2008a,FCEC2011a}.

The key point is {\it not} to have prices with privileged status from any source.  This is doubly true when those privileges can increase.   Essentially we advocate a skepticism towards making pricing systematically institutional with legal and regulatory privileges.  This is a direct consequence of privileging CCPs in terms of capital over, for example, bilateral collateralization\footnote{There is a similar issue w.r.t. tri-party repo netting versus bilateral repo netting that also pushes towards CCPs.}.  The primary function of CCPs, removal of credit risk transmission,  is accomplished by collateralization with initial margins.  It is not clear that a CCP is a requirement.  If the removal of credit risk transmission is desired then we advocate equal regulatory treatment of all equivalent collateralization arrangements.  Collateralization itself does not require institutionally privileged prices.

\bibliographystyle{alpha}
\bibliography{kenyon_general}
\end{document}